\documentclass[letterpaper, 10 pt, conference]{ieeeconf}  

\IEEEoverridecommandlockouts                              
\usepackage{cite}
\usepackage{amsmath,amssymb,amsfonts}
\usepackage{graphicx}
\usepackage{textcomp}
\usepackage{xcolor}
\usepackage{algorithm}
\usepackage{algpseudocode}
\usepackage{multirow}
\usepackage{algorithm}
\usepackage{algpseudocode}

\newtheorem{assumption}{Assumption}
\newtheorem{definition}{Definition}
\newtheorem{lemma}{Lemma}
\newtheorem{theorem}{Theorem}

\newtheorem{proposition}{Proposition}
\newtheorem{property}{Property}

\title{\LARGE \bf
Zonotope-Based Active Exposure of Stealthy Deception Attacks in Sensor-Fusion Systems*
}

\author{Meiqi Tian$^{1}$, Shuo Li$^{1}$ and Bingzhuo Zhong$^{1}$
\thanks{*This work was supported by Guangzhou-HKUST(GZ) Joint Funding Program (Grant No. 2025A03J4493), Education Bureau of Guangzhou Municipality, Guangdong Provincial Project 2024QN11X053, and the Youth S\&T Talent Support Programme of GDSTA (SKXRC2025468). (Corresponding author: Bingzhuo Zhong.)}
\thanks{$^{1}$Meiqi Tian, Shuo Li and Bingzhuo Zhong are with the Thrust of Artificial Intelligence, The Hong Kong University of Science and Technology (Guangzhou), Guangzhou 511400, China. (e-mail:\{mtian837, sli430\}@connect.hkust-gz.edu.cn, bingzhuoz@hkust-gz.edu.cn)}
}

\begin{document}

\maketitle
\thispagestyle{empty}
\pagestyle{empty}

\begin{abstract}

This paper investigates the stealthy attack detection for sensor-fusion cyber-physical systems with unknown-but-bounded noises through the control channel. 
The detection framework is particularly applicable to sensor-fusion scenarios in which multiple suspicious sensors contributing to the fused estimate may be compromised simultaneously.
First, we construct an admissible output set using secure sensors and an attack output set for each attack hypothesis.
Then, we introduce a receding-horizon optimization framework to design exposure inputs, namely bounded auxiliary control perturbations injected through the control channel, so as to enlarge the separation between the admissible output set and the attack output sets according to the separation tendency.
A sufficient detection condition is further derived, showing that set separation guarantees detectability of the compromised sensors. 
Moreover, an offline exposure budget guidance is developed to support budget selection before online exposure starts. 
Simulations on a UAV navigation system under stealthy GNSS and LiDAR attacks validate the proposed method.

\end{abstract}

\section{INTRODUCTION}
Sensor fusion has become an imperative component in modern cyber-physical systems (CPSs), as it enables accurate and consistent state estimation by integrating heterogeneous sensing information. 
However, the reliance on sensor feedback and vulnerable communication channels also enlarges the attack surface of such systems. 
Existing studies have shown that intelligent adversaries can construct stealthy deception attacks using data driven, dynamic programming, and learning-based strategies~\cite{attackmethod1,attackmethod2,attackmethod3}. By manipulating one or multiple sensing channels while remaining undetectable.

Secure control schemes have gained prominence in deception attack defense. 
A secure model predictive control method and a fault tolerant control barrier function for safety control synthesis are studied in~\cite{wang2020security} and~\cite{CBF2024}, respectively.
The reinforcement learning-based secure tracking and Stackelberg game-based optimal secure control are studied in~\cite{Wu2023} and~\cite{Xiong2025}, respectively.
However, they mainly tolerate attack effects rather than actively reveal stealthy attacks, allowing malicious behavior to remain hidden.

To address this issue, active attack detection has attracted increasing attention.
Active detection methods identify counterfeit measurements by injecting specially designed perturbations and verifying the consistency between the returned measurements and the injected signals.
Representative schemes include packet modification and dynamic watermarking~\cite{Pang2021, watermarking1, watermarking2}.
These methods provide strong detection capabilities, but they often rely on a specific processing architecture, which involves a trusted sender and receiver. 

Another line of work detects attacks through cross-modal or state-consistency checks, including IMU–GNSS spoofing detection~\cite{IMU2022}, state-inconsistency-based perception attack detection~\cite{Inconsistency2025}, and LiDAR–camera consistency checking~\cite{lidar}. However, these methods are typically tailored to specific sensor pairings and mainly consider a single compromised sensor, leaving simultaneous multi-sensor attacks less explored.

Zonotopes provide a flexible and computationally efficient tool for set representation, offering distinct advantages for reachability analysis.
Zonotope-based methods have also been investigated for replay attack detection, attack estimation, and false data injection detection~\cite{zonotope2, zonotope3, zonotope1}.
However, these studies are mainly built on passive detection mechanisms, which may be ineffective against well-designed stealthy deception strategies.
Moreover, they do not address the structural challenges of sensor-fusion systems.

Motivated by the limitations discussed above, this paper investigates the exposure of stealthy deception attacks in sensor-fusion systems, where multiple suspicious sensors may be compromised simultaneously.
The main contributions are summarized as follows.
\begin{itemize}
    \item We propose a zonotope-based active exposure framework for stealthy deception attacks. Bounded control perturbations are designed to enlarge the separation between defender-side admissible output sets and hypothesis-dependent attack output sets, enabling inconsistent hypotheses to be excluded without imposing special requirements on the sensing or communication architecture.
    \item We establish a sufficient set-separation condition for attack detection and develop a receding-horizon exposure strategy with hypothesis updating. In addition, offline lower and sufficient budget thresholds are derived to guide the selection of the exposure-input magnitude before online exposure.
\end{itemize}

\textit{Notation:}
Let $\mathbb{R}$, $\mathbb{R}^n$ and $\mathbb{R}^{n \times m}$ be the set of real numbers, $n$-dimensional real vectors, and $n \times m$ real matrices, respectively.
Let $\mathbb{N}_+$ and $\mathbb{R}_+$ denote the set of positive natural numbers and positive real numbers, respectively.
For vectors \(x,y\in\mathbb{R}^n\), the notation \(|x|\preceq y\) means that \(|x_i|\le y_i\) for all \(i=1,\dots,n\).
$(\cdot)_+$ represents the positive part operator satisfying $(n)_+ = \max\{n,0\}$, and $\operatorname{col}(\cdot)$ denotes the column-stacking operator.
 

\section{Preliminaries and Problem Formulation}\label{sec: 2}
\subsection{Preliminaries}
\begin{definition}[\cite{bib3}]
A $p$-order zonotope $\mathcal{Z} \subseteq \mathbb{R}^n$ is defined as
\[
\mathcal{Z}=\langle c,H\rangle
:=\{\,c+Hz \mid z\in[-1,1]^p\,\},
\]
where $c \in \mathbb{R}^n$ is the center of the zonotope, and
$H \in \mathbb{R}^{n \times p}$ is the generator matrix of $\mathcal{Z}$.
\end{definition}

\begin{property}[\cite{bib4}]
Given two zonotopes $\mathcal{Z}_1=\langle c_1,H_1\rangle$ and
$\mathcal{Z}_2=\langle c_2,H_2\rangle$, their Minkowski sum and linear map satisfy
\begin{align*}
    \langle c_1,H_1\rangle\oplus\langle c_2,H_2\rangle
&=\langle c_1+c_2,\ [H_1 \; H_2]\rangle,\\
K\langle c_1,H_1\rangle &= \langle Kc_1,KH_1\rangle,
\end{align*}
where $K$ is a matrix of compatible dimension.
\end{property}


\begin{figure}
    \centering
    \includegraphics[width=0.8\linewidth]{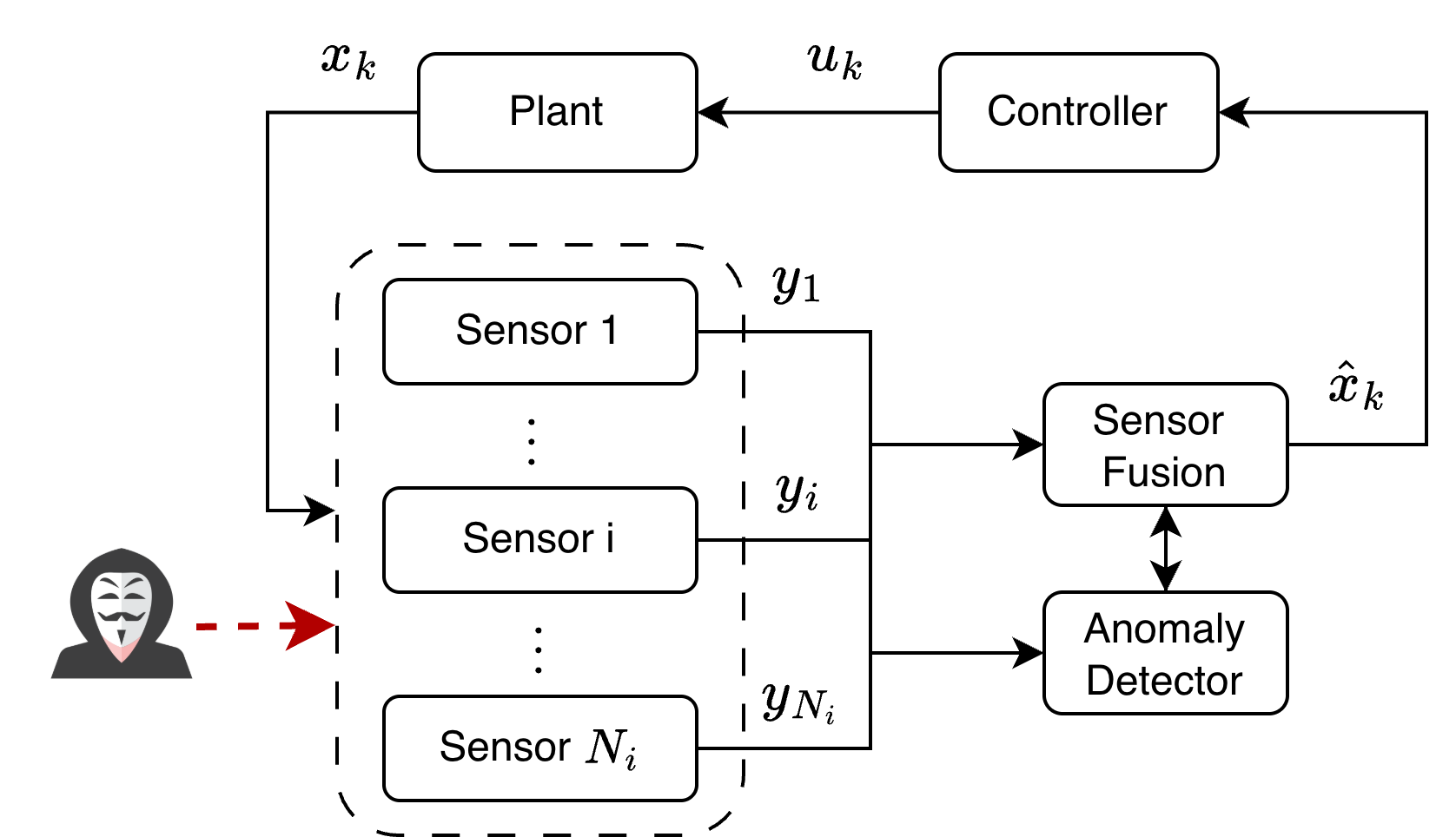}
    \caption{General architecture of sensor-fusion system under deception attack.}
    \label{fig:placeholder}
\end{figure}

\subsection{System Model}
In this paper, we consider the following discrete-time linear time-invariant system:
\begin{align}\label{eq: system model}
\begin{cases}
    x(k+1) =Ax(k)+Bu(k)+w(k),\\
    y_{i}(k) = C_ix(k)+ v_{i}(k)+ a_{i}(k),
\end{cases}
\end{align}
where \(x(k)\in \mathbb{R}^{n_x}\) denotes the system state,
\(u(k)\in \mathbb{R}^{n_u}\) denotes the control input,
\(A \in \mathbb{R}^{n_x \times n_x}\), \(B \in \mathbb{R}^{n_x\times n_u}\), and
\(C_i \in \mathbb{R}^{n_{y_i}\times n_x}\) are matrices of compatible dimensions,
\(w(k) \in \mathbb{R}^{n_x}\) denotes the process noise,
\(v_i(k) \in \mathbb{R}^{n_{y_i}}\) denotes the measurement noise of the \(i\)-th sensor,
and \(a_i(k) \in \mathbb{R}^{n_{y_i}}\) denotes the attack vector injected into the \(i\)-th sensor channel.

\begin{assumption}
$w(k)$ and $v_i(k)$ belong to the zonotopes
$\mathcal{W} = \langle w^c, H_w \rangle$ and
$\mathcal{V}_i = \langle v^{i,c}, H_{i,v}\rangle$, respectively, where
\(
w^c\in\mathbb{R}^{n_x},
v^{i,c}\in \mathbb{R}^{n_{y_i}},
H_w \in \mathbb{R}^{n_x\times p_w},
H_{i,v}\in\mathbb{R}^{n_{y_i}\times p_{v_i}}.
\)
\end{assumption}

A feedback controller is employed as
\begin{equation}\label{eq: controller}
u(k) = K(\bar x(k)-\hat x(k)),
\end{equation}
where \(K \in \mathbb{R}^{n_u\times n_x}\) is a fixed feedback gain,
\(\bar x(k) \in \mathbb{R}^{n_x}\) is the desired state, and
\(\hat x(k) \in \mathbb{R}^{n_x}\) is the fused state estimate, which is generated by a sensor fusion module whose specific architecture is not restricted in this paper.
The overall architecture is illustrated in Fig.~\ref{fig:placeholder}.

To monitor the consistency of sensor measurements, the system utilizes an anomaly detector associated with the sensor-fusion process.
The detector is kept generic and may represent any standard residual-, innovation-, or consistency-checking thresholding mechanism~\cite{manandhar2014detection, jin2024detection, dasgupta2022sensor}.
Let \(q(k)\) denote the corresponding detection statistic. An alarm is triggered at time step \(k\) whenever
\begin{align}
    q(k) > \tau(k), \label{threshold}
\end{align}
where \(\tau(k)\) is a prescribed threshold.

Based on the detector outputs and prior system knowledge, the subset of sensors regarded as suspicious are collected into \textit{suspicious sensor set} $\mathcal{A}$.
The remaining sensors form the \textit{secure sensor set} $\mathcal{S}$.
Since each suspicious sensor can be either compromised or uncompromised, there are $2^{|\mathcal{A}|}$ possible attack hypotheses.
Let $\mathcal{H}:=\{1,\ldots,2^{|\mathcal{A}|}\}$ denote the index set of all attack hypotheses.
For each $h\in\mathcal{H}$, let $\mathcal{F}(h)\subseteq \mathcal{A}$ denote the set of sensors assumed to be compromised under hypothesis $h$.
Hence, if attack hypothesis $h$ occurs, then the outputs of the sensors in $\mathcal{F}(h)$ can be arbitrarily manipulated by the attacker. We assume that all attack hypotheses are available, while the true attack hypothesis remains unknown.

\subsection{Attacker and Defender Model}
We consider an omniscient attacker whose capabilities are specified as follows.

\begin{assumption}\label{asm:attacker}
The attacker is modeled as a white-box adversary with perfect knowledge of:
(i) the system dynamics in \eqref{eq: system model}, including $A$, $B$, and the uncertainty sets $\mathcal{W}$ and $\mathcal{V}_i$;
(ii) the feedback controller as in~\eqref{eq: controller}; and
(iii) the threshold $\tau(k)$ of the anomaly detector as defined in~\eqref{threshold}.
\end{assumption}

Under Assumption~\ref{asm:attacker}, the attacker can maintain an internal estimation-and-deception process, which enables the attacker not only to emulate the nominal closed-loop behavior and remain stealthy, but also to strategically choose attack actions serving its attack objective.

For each attack hypothesis $h\in\mathcal H$, we consider the following attacker model:
\begin{align}\label{eq: attacker dynamics}
\begin{cases}
\hat x_h^{a-}(k+1)= A\hat x_h^{a}(k) + Bu_h^a(k)+ w_h^a(k),\\
\hat x_h^{a}(k+1)= \hat x_h^{a-}(k+1) + \Delta x_h^a(k),
\end{cases}
\end{align}
where \(A\) and \(B\) are the same matrices as in \eqref{eq: system model},
\(w_h^a(k)\in\mathbb R^{n_x}\) denotes the estimated process noise by the attacker,
\(u_h^a(k)\) denotes the control input reconstructed according to the control law~\eqref{eq: controller},
\(\Delta x_h^a(k)\) denotes the deviation induced by the deception attack, and
\(\hat x_h^{a-}(k+1)\) and \(\hat x_h^{a}(k+1)\) denote the attacker’s state estimate before and after attack injection, respectively.

The following definition formalizes stealthiness in terms of the tolerance region allowed by the anomaly detector.

\begin{definition}\label{def: stealthy attack}
Consider the attacker model~\eqref{eq: attacker dynamics}.
An attack under hypothesis $h$ is said to be stealthy if its induced deviation $\Delta x_h^a(k)$ satisfies
\begin{equation}\label{eq: stealthy}
|\Delta x_h^a(k)| \preceq T(k),
\end{equation}
where $T(k)\in\mathbb R^{n_x}_{+}$ is a vector derived from the threshold of the anomaly detector.
\end{definition}

We next construct a defender model based on the secure sensor set as
\begin{align}\label{eq: defender dynamics}
\begin{cases}
x^s(k+1) = Ax^s(k) + Bu^s(k) + w(k),\\
y_{\mathcal{S}}(k) = C_{\mathcal{S}}x^s(k) + v_{\mathcal{S}}(k),
\end{cases}
\end{align}
where \(
y_{\mathcal{S}}(k)=\operatorname{col}(y_i(k))_{i\in\mathcal S},\;
C_{\mathcal{S}}=\operatorname{col}(C_i)_{i\in\mathcal S},\) and \(
v_{\mathcal{S}}(k)=\operatorname{col}(v_i(k))_{i\in\mathcal S}
\).

For each secure sensor \(i\in\mathcal S\), suppose that the measurement noise satisfies
\(v_i(k)\in \mathcal V_i=\langle v^{i,c},H_{i,v}\rangle\).
Then the stacked secure sensor noise satisfies
\[
v_{\mathcal S}(k)\in \mathcal V_{\mathcal S}=\langle v_{\mathcal S}^c,H_{\mathcal S,v}\rangle,
\]
where \(v_{\mathcal S}^c=\operatorname{col}(v^{i,c})_{i\in\mathcal S},\) and 
\(H_{\mathcal S,v}=\operatorname{blkdiag}(H_{i,v})_{i\in\mathcal S}\).

\subsection{Problem of Interest}
Given the secure sensor set \(\mathcal S\), the suspicious sensor set \(\mathcal A\), the attacker model ~\eqref{eq: attacker dynamics}, and the defender model~\eqref{eq: defender dynamics}, our goal is to identify the true attacked sensors in a prescribed horizon.
Technically, we seek to design a bounded exposure input sequence that can exclude inconsistent attack hypotheses and make the stealthy attack detectable.

\section{Set Construction and Detection Criterion}\label{sec: 3}
In this section, we first construct the secure state set by combining the predicted state set with the secure sensor measurements, and then establish a sensor-level detection criterion. 
We further introduce hypothesis-level output sets for exposure input design in Section~\ref{sec: Exposure Input Design}.

\subsection{Secure State Set Construction}
To characterize the set of system states consistent with the secure sensor information, we construct a secure state set based on the following sets.
\begin{definition}[\cite{bib3} (\textit{Predicted State Set})]
For the defender model \eqref{eq: defender dynamics}, suppose that the state $x(k-1)$ belongs to a zonotope $\mathcal{X}(k-1) = \langle c(k-1),\ H(k-1) \rangle \subseteq \mathbb{R}^{n_x}$, $k \in \mathbb{N}_+$. 
The set of all possible states $x(k)$ is defined as the predicted state set $\mathcal{X}(k|k-1)$, i.e.,
    \begin{align*}
    \mathcal{X}&(k|k-1) = \\
    &\big\{ x(k) \in \mathbb{R}^{n_x} \mid x(k) \in A\mathcal{X}(k-1) \oplus \{Bu(k-1)\} \oplus \mathcal{W} \big\}.
    \end{align*}
\end{definition}
\begin{definition}[\cite{bib3} (\textit{Measurement State Set})]\label{def: Measurement State Set}
For the defender model \eqref{eq: defender dynamics},  define the set of all possible states $x(k)$ consistent with the measurement of the $i$-th sensor as the $i$-th measurement state set $\mathcal{X}_{y_i}(k)$, i.e.,
\begin{equation}\label{eq: Measurement State Set eq}
    \mathcal{X}_{y_i}(k) = \big\{ x(k) \in \mathbb{R}^{n_x} \mid (y_i(k) - C_i x(k)) \in \mathcal{V}_{i} \big\}, \quad k \in \mathbb{N}_+.\nonumber
\end{equation}
\end{definition}

From Definition~\ref{def: Measurement State Set}, we define the set of all possible states consistent with the secure sensor measurements as the \textit{secure measurement state set} \(\mathcal{X}_{y_{\mathcal S}}(k)\), i.e.,
\begin{equation}
\mathcal{X}_{y_{\mathcal S}}(k)
=
\left\{
x(k)\in\mathbb R^{n_x}\;\middle|\;
y_{\mathcal S}(k)-C_{\mathcal S}x(k)\in\mathcal V_{\mathcal S}
\right\},
\; k\in\mathbb N_+. \nonumber
\end{equation}

Based on the predicted state set \(\mathcal X(k|k-1)\) and secure measurement state set \(\mathcal X_{y_{\mathcal S}}(k)\), we adopt the construction method in~\cite{Zhao} to build the \textit{secure state set} \(\mathcal X_{\mathcal S}(k)\) as follows.

\begin{lemma}\label{lemma: secure state set}
Given the predicted state set \(\mathcal X(k|k-1)=\langle c(k-1),H(k-1)\rangle\), the intersection of \(\mathcal X(k|k-1)\) and \(\mathcal X_{y_{\mathcal S}}(k)\) can be over-approximated by the zonotope
\[
\mathcal X_{\mathcal S}(k)=\langle c_{\mathcal S}(k),H_{\mathcal S}(k)\rangle,
\]
whose center and generator matrix are recursively given by
\begin{equation}
\begin{aligned}
c_{\mathcal S}(k)
={}&Ac(k-1)+Bu(k-1)+L_{\mathcal S}(k) \\
&(y_{\mathcal S}(k)-C_{\mathcal S}Ac(k-1)-C_{\mathcal S}Bu(k-1)),
\end{aligned}
\end{equation}
and
\begin{equation}
\begin{aligned}
H_{\mathcal S}(k)
=
\Big[
&(A-L_{\mathcal S}(k)C_{\mathcal S}A)H(k-1), \\
&(I-L_{\mathcal S}(k)C_{\mathcal S})H_w,\;
-L_{\mathcal S}(k)H_{\mathcal S,v}
\Big],
\end{aligned}
\end{equation}
with
\[
L_{\mathcal S}(k)=
\big[A\Phi(k-1)A^\top+\Phi_w\big]C_{\mathcal S}^\top
\Theta_{\mathcal S}^{-1}(k-1),
\]
where 
\(\Phi(k-1)=H(k-1)H^\mathrm{T}(k-1), 
\Phi_w=H_wH_w^\mathrm{T},
\Phi_{\mathcal{S},v}=H_{\mathcal{S},v}H_{\mathcal{S},v}^\top,
\)
and \(\Theta^{-1}(k-1)=C_{\mathcal{S}}\big[A\Phi(k-1)A^\mathrm{T}+\Phi_w\big]C_{\mathcal{S}}^\mathrm{T}+\Phi_{\mathcal{S},v}.\)
Here, \(L_{\mathcal S}(k)\) is a correction matrix chosen to minimize the Frobenius norm of the generator matrix \(H_{\mathcal S}(k)\), i.e., \(\|H_{\mathcal S}(k)\|_F^2=\operatorname{tr}(H^\top_{\mathcal S}(k) H_{\mathcal S}(k))\).
\end{lemma}

\subsection{Detection Criterion}
The following theorem establishes a sufficient detection condition for the proposed exposure framework.

\begin{theorem}\label{thm:detection_condition}
Consider the defender model \eqref{eq: defender dynamics} and the secure
state set \(\mathcal X_{\mathcal S}(k)\) constructed by Lemma~\ref{lemma: secure state set}.
For each suspicious sensor
\(i\in\mathcal A\), define the sensor-level admissible output set as
\begin{equation}
\mathcal Y_i(k):=C_i\mathcal X_{\mathcal S}(k)\oplus \mathcal V_i .
\label{eq:sensor_admissible_set}
\end{equation}
Then, an attack on sensor \(i\) is detected at time step \(k\) if
\begin{equation}
y_i(k)\notin \mathcal Y_i(k).
\label{eq:sensor_detection_condition}
\end{equation}
\end{theorem}

This sensor-level criterion determines whether the output of an individual suspicious sensor is inconsistent with its admissible output set. 
By contrast, the exposure input is designed at the hypothesis level by simultaneously considering candidate attacked sensor subsets, so as to separate the defender-side and attacker-side output sets, and thereby facilitate the exclusion of inconsistent hypotheses.

\subsection{Hypothesis-Level Output Set Construction}
To support exposure input design, we next define the hypothesis-level admissible output set and attack output set.

Based on the secure state set ${\mathcal{X}}_\mathcal{S}(k)$, for each attack hypothesis \(h\in\mathcal H\), the hypothesis-level admissible output set is constructed as
\begin{equation}\label{eq:hypothesis-level admissible}
    \mathcal Y_h(k)=C_h \mathcal{X}_{\mathcal S}(k)\oplus \mathcal V_h, \forall h \in\mathcal{H}.
\end{equation}
Here \(C_h=\operatorname{col}(C_i)_{i\in \mathcal{F}(h)}\) and \(\mathcal{V}_h=\langle v_h^c,\, H_{h,v}\rangle\) are the stacked measurement set and stacked measurement noise set of hypothesis \(h\),
where \(v_h^c=\operatorname{col}(v^{i,c})_{i\in \mathcal{F}(h)},\) and \(
H_{h,v}=\operatorname{blkdiag}(H_{i,v})_{i\in \mathcal{F}(h)}.\)

Let \(\hat{\mathcal X}_h(k)\) denote the attacker-side reachable state set.
As the attacker is assumed to know and reconstruct the same feedback law as the defender, the corresponding attack reachable set \(\hat{\mathcal X}_h(k)\) evolves as
\begin{equation}
\hat{\mathcal X}_h(k+1) = (A-BK)\hat{\mathcal X}_h(k)
\oplus \{BK\bar x(k)\}
\oplus \mathcal W_h
\oplus \Delta \mathcal X_h^a(k),\nonumber
\label{eq: attack reachable set}
\end{equation}
where \(\mathcal W_h\subseteq\mathbb R^{n_x}\) denotes the attacker-side process noise set, and \(\Delta\mathcal X_h^a(k)\subseteq\mathbb R^{n_x}\) denotes the set of admissible attack-induced deviations under hypothesis \(h\).

The corresponding attack output set is then given by
\begin{equation}
\hat{\mathcal Y}_h(k)
=
C_h\hat{\mathcal X}_h(k)\oplus \mathcal V_h.
\end{equation}

\begin{figure}
    \centering
    \includegraphics[width=0.9\linewidth]{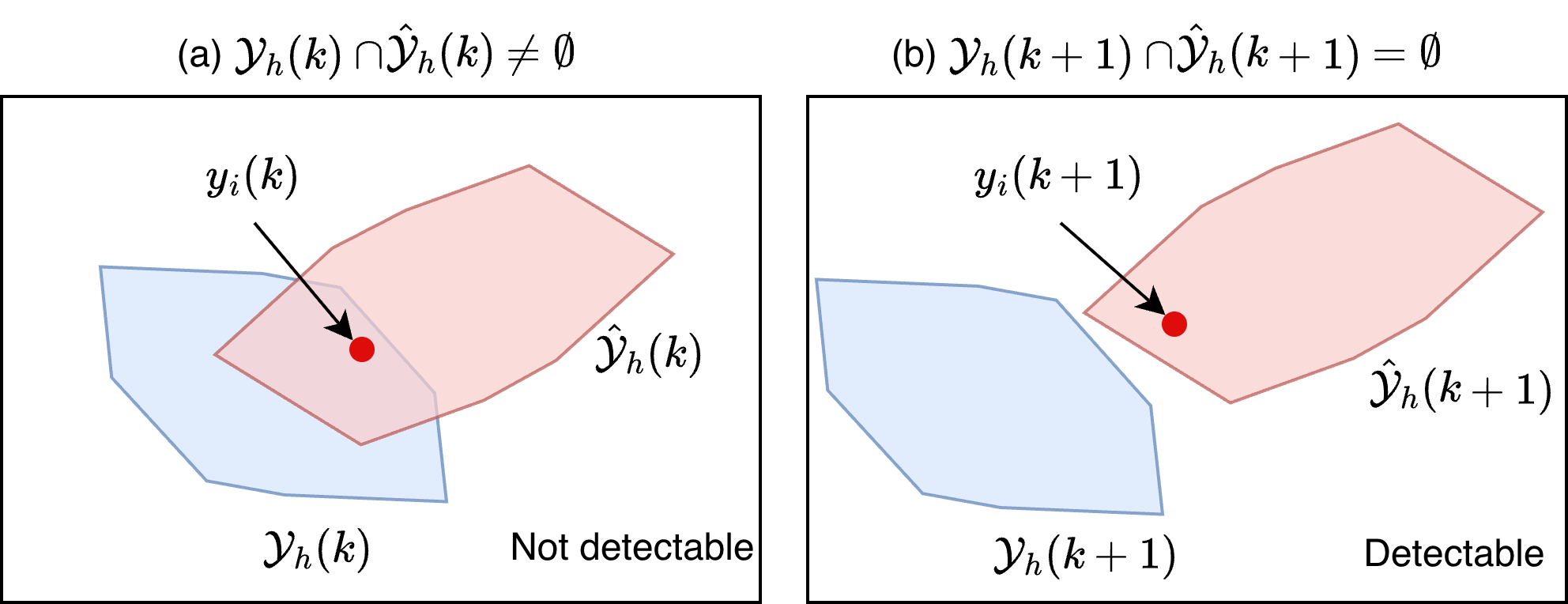}
    \caption{Illustration of zonotope separation-based attack exposure. The blue zonotope $\mathcal{Y}_h(k)$ denotes the admissible output set, while the red zonotope $\hat{\mathcal{Y}}_h(k)$ denotes the attack output set. The red dot $y_i(k)$ represents the actual received sensor output.}
    \label{fig: set separation}
\end{figure}

The set \(\mathcal Y_h(k)\) collects all stacked outputs that remain consistent
with the secure state set, while
\(\hat{\mathcal Y}_h(k)\) represents the stacked outputs generated by the
corresponding attack reachable set under hypothesis \(h\). 
Hence, enlarging the separation between
\(\mathcal Y_h(k)\) and \(\hat{\mathcal Y}_h(k)\) makes the attacked sensors
under hypothesis \(h\) more likely to violate their sensor-level admissible
output sets.
Fig.~\ref{fig: set separation} provides an intuitive illustration of this zonotope-separation-based attack exposure.

\section{Exposure Input Design}\label{sec: Exposure Input Design}
To reveal stealthy deception attacks, exposure inputs are injected into the nominal control channel to enlarge the separation between the admissible output set and the attack output sets, thereby making stealthy attacks detectable. 
As the actually compromised sensors are detected, the attack hypothesis set can be narrowed down.

In this section, we first present an online exposure input generation method, and then develop an offline budget guidance to provide a practical reference range for selecting the exposure budget, so that effective attack exposure can be achieved without introducing unnecessarily large control perturbations.

The composite control input is described as
\begin{equation}\label{eq: input_decomp}
u^s(k)= u^*(k) + d(k),\qquad \|d(k)\|_\infty \le \bar u,
\end{equation}
where $u^*(k) \in \mathbb{R}^{n_u}$ is the nominal tracking input provided by the control law~\eqref{eq: controller}, and $d(k) \in \mathbb{R}^{n_u}$ is the exposure input. 
The scalar $\bar u \in \mathbb{R}_+$ represents the available exposure budget, and its selection guidance will be given later.

To quantify the separation level between two zonotopes, we adopt the following definition.

\begin{definition}[\cite{separationTendency} (\textit{Separation Tendency})]\label{def: separation tendency}
Given two zonotopes \(\mathcal Z_1=\langle c_1,H_1\rangle\) and \(\mathcal Z_2=\langle c_2,H_2\rangle\), their separation tendency is defined as the optimal value of
\begin{equation}
\begin{aligned}
\hat\delta(\mathcal Z_1,\mathcal Z_2)
&=\;\min_{\delta,\xi_1,\xi_2}\ \delta\\
\text{s.t.}\quad&c_1+H_1\xi_1=c_2+H_2\xi_2,\\
&\|\xi_1\|_\infty\le\delta,\quad \|\xi_2\|_\infty\le\delta.
\end{aligned}
\end{equation}
\end{definition}

Intuitively, \(\hat\delta(\mathcal Z_1,\mathcal Z_2)\) is the smallest common scaling factor on the generator coefficients that makes the two zonotopes intersect. Hence, \(\hat\delta(\mathcal Z_1,\mathcal Z_2)>1\) indicates that the two zonotopes are already separated, i.e., \(\mathcal Z_1\cap\mathcal Z_2=\emptyset\), whereas \(\hat\delta(\mathcal Z_1,\mathcal Z_2)\le 1\) indicates that overlap is still possible under the original uncertainty bounds, i.e., \(\mathcal Z_1\cap\mathcal Z_2 \neq \emptyset.\)

According to Definition~\ref{def: separation tendency} and Theorem~\ref{thm:detection_condition}, our objective is to find an exposure input sequence $[d(t),\cdots, d(t+N-1)]$ such that 
\begin{equation}\label{eq: separation goal}
    \exists k \in [t, t+N],\mathcal{Y}_h(k) \cap \hat{\mathcal{Y}}_h(k) = \emptyset, \forall h \in \mathcal{H},
\end{equation}
which is equivalent to
\begin{equation}\label{eq: separation tendency greater than 1}
\exists k \in [t, t+N],\;\hat \delta(\mathcal{Y}_h(k), \hat {\mathcal{Y}}_h(k)) > 1, \; \forall h \in \mathcal{H}.
\end{equation}
Therefore, the exposure input design can be viewed as a process of enlarging the hypothesis-level separation tendency.
To ensure that the injected exposure inputs are visible at the corresponding hypothesis-level outputs over the exposure horizon, the following assumption is required.

\begin{assumption}\label{ass:output_visibility}
For each attack hypothesis \(h\in\mathcal H\), there exists an integer
\(r_h\in\{0,\dots,N-1\}\), such that 
\(
\|C_hA^{r_h}B\|_{\infty}> 0.
\)
\end{assumption}

Under Assumption~\ref{ass:output_visibility}, the injected exposure input has an effect on the output associated with every attack hypothesis within the prescribed horizon. This allows us to formulate an online design strategy that enlarges the separation tendency over all attack hypotheses.

\subsection{Online Exposure Input Generation}
To achieve~\eqref{eq: separation tendency greater than 1}, we adopt a receding-horizon strategy
to design the exposure input, which is formulated as
\begin{subequations}\label{eq:online_opt_output}
\begin{align}
\max_{d(k),\,\gamma(k+1)}\quad
& \sum_{h\in\mathcal H(k)} w_h\,\hat \delta(\mathcal{Y}_h(k+1), \hat{\mathcal{Y}}_h(k+1))
\label{eq: opt1}\\
\text{s.t.}\quad
& \hat \delta(\mathcal{Y}_h(k+1), \hat{\mathcal{Y}}_h(k+1))\ge \gamma(k+1),
\notag\\
& \forall h\in\mathcal H(k),
\label{eq: opt2}\\
& \gamma(k+1)\ge \gamma(k)+\varepsilon,
\label{eq: opt3}\\
& \|d(k)\|_\infty\le \bar u.
\label{eq: opt4}
\end{align}
\end{subequations}
where $\mathcal H(k)$ denotes the remaining attack hypotheses at time step $k$, $w_h\in\mathbb R_{+}$ denotes the weight associated with the
attack hypothesis $h$, and $\varepsilon>0$ specifies the desired
per-step increase of the worst-case separation level.

\subsubsection*{Hypothesis update}
Sensor-level detection is performed continuously during the exposure phase for each suspicious sensor. 
Once sensor \(i\) is identified as attacked according to \eqref{eq:sensor_detection_condition}, all attack hypotheses that do not include sensor \(i\) are eliminated from the remaining hypothesis set. Accordingly, the hypothesis set is updated as
\begin{equation}
\mathcal H(k+1)=\{\,h\in\mathcal H(k):\ i\in \mathcal F(h)\,\}.
\end{equation}

\subsubsection*{Stopping Rule}
The exposure phase is terminated when one of the following conditions is met:
\begin{enumerate}[label=(\roman*)]
    \item the remaining hypothesis set reduces to a singleton, i.e., \(\mathcal{H}(k)=\{h^\star\}\), and the corresponding separation tendency satisfies \(\delta_{h^\star}(k)>1\);
    \item the remaining hypothesis set contains more than one element, i.e., \(|\mathcal{H}(k)|>1\), but the corresponding separation tendency satisfies \(\delta_h(k)>1\) for all \(h\in \mathcal{H}(k)\);
    \item the prescribed horizon is exhausted, i.e., \(k=t+N\).
\end{enumerate}

Conditions (i) and (ii) indicate that the exposure objective has been achieved for the current remaining hypothesis set. 
In particular, condition (i) implies that the attacked sensor subset has been uniquely identified. 
Condition (ii) implies that the sensors associated with  $\mathcal{H}(k)$ are not attacked.
Condition (iii) means that the horizon is exhausted, and the sensors associated with $\mathcal{H}(k)$ remain undetectable. 

\subsection{Offline Budget Guidance}
Before starting the exposure phase, it is desirable to obtain offline guidance for selecting the exposure budget $\bar u$ used in \eqref{eq: opt4}.
The purpose of this guidance is to provide a practical reference range for choosing $\bar u$, so that the exposure input is sufficiently effective for attack exposure while introducing limited impact on nominal control performance.
Since directly quantifying the relationship between the separation objective~\eqref{eq: separation tendency greater than 1} and $\bar u$ over a horizon would lead to a nested optimization problem, we introduce predicted sets and a surrogate separation tendency based on the center mismatch and the uncertainty radius.

Assume that the exposure phase starts at time \(t_0\). We first define the
predicted secure set as
\[
\bar X_\mathcal{S}(k)=\langle \bar c_\mathcal{S}(k),\bar H_\mathcal{S}(k)\rangle,
\]
initialized by \(\bar {\mathcal X}_\mathcal{S}(t_0)={\mathcal X}_\mathcal{S}(t_0)\). Its center and generator are propagated according to
\begin{align}
\bar c_\mathcal{S}(k+1) &= A\bar c_\mathcal{S}(k)+B\big(u^n(k)+d(k)\big), \label{eq:barcS}\\
\bar H_\mathcal{S}(k+1) &= \big[A\bar H_\mathcal{S}(k)\;\; H_w\big]. \label{eq:barHS}
\end{align}
Here, $u^n(k)$ is generated by \(u^n(k)=K\big(\bar x(k)-x^n(k)\big)\),
where $x^n(k)$ is the predicted nominal center state propagated by \(x^n(k+1)=Ax^n(k)+Bu^n(k)\).

Let \(\bar c_\mathcal{S}^0(k)\) denote the nominal
predicted secure center generated with \(d(k)\equiv 0\), i.e., \(\bar c_\mathcal{S}^0(k+1)=A\bar c_\mathcal{S}^0(k)+B u^n(k)\), \(\bar c_\mathcal{S}^0(t_0)=c_\mathcal{S}(t_0)\).
For each attack hypothesis \(h\in \mathcal{H}\) and \(\ell=\{1,\ldots,N\}\), the predicted admissible
output set and the attack output set are defined as
\begin{align}
\bar Y_h(t_0+\ell)
&:= C_h\bar {\mathcal X}_\mathcal{S}(t_0+\ell)\oplus V_h
   = \langle \bar c_h(\ell), \bar H_h(\ell)\rangle, \label{eq:22}\\
\hat Y_h(t_0+\ell)
&:= C_h\hat {\mathcal X}_h(t_0+\ell)\oplus V_h
   = \langle \hat c_h(\ell), \hat H_h(\ell)\rangle, \label{eq:23}
\end{align}
where
\[
\bar c_h(\ell)=C_h\bar c_\mathcal{S}(t_0+\ell)+v_h^c,\;
\bar H_h(\ell)=[C_h\bar H_\mathcal{S}(t_0+\ell)\;\; H_{h,v}],
\]
\[
\hat c_h(\ell)=C_hc_h(t_0+\ell)+v_h^c,\;
\hat H_h(\ell)=[C_hH_h(t_0+\ell) \;\; H_{h,v}].
\]

Based on these predicted output sets, we introduce the surrogate separation tendency as
\begin{equation}\label{eq: surrogate separation tendency definition}
\bar \delta_h(\ell)=
\frac{\|\bar c_h(\ell)-\hat c_h(\ell)\|_\infty}{\rho_h(\ell)},
\end{equation}
where
\[
\rho_h(\ell)=
\left\|[ C_h\bar H_\mathcal{S}(t_0+\ell) \;\; C_hH_h(t_0+\ell) \;\; H_{h,v} \;\; H_{h,v}]
\right\|_\infty .
\]
Here, \(\rho_h(\ell)\) characterizes the uncertainty radius
of the two output sets. 
\(\bar \delta_h(\ell)\) measures
the separation level from the perspective of the center mismatch
relative to the aggregated uncertainty.

To reveal how the exposure budget enters this surrogate metric, we define the nominal center mismatch as
\begin{equation}\label{eq: e_h^0}
e_h^0(\ell)=C_h\bar c_\mathcal{S}^0(t_0+\ell)-C_hc_h(t_0+\ell),
\end{equation}
the stacked exposure-input sequence over \([t_0,t_0+\ell-1]\) as
\begin{equation}
    d_\ell =
\begin{bmatrix}
d^\top(t_0) & d^\top(t_0+1) & \cdots & d^\top(t_0+\ell-1)
\end{bmatrix}^\top,
\end{equation}
and the finite-horizon input-propagation matrix as
\begin{equation}
    M_\ell =
[A^{\ell-1}B \;\; A^{\ell-2}B \;\; \cdots \;\; B].
\end{equation}

By repeatedly expanding \eqref{eq:barcS}, one has
\begin{equation}
\bar c_h(\ell)-\hat c_h(\ell)
=
e_h^0(\ell)+C_hM_\ell d_\ell.
\label{eq:25}
\end{equation}
Hence, the surrogate separation tendency depends on the
available budget \(\bar u\) through the stacked exposure input
sequence \(d_\ell\).

\begin{proposition}\label{prop:surrogate_sep}
For any attack hypothesis \(h\in \mathcal{H}\) and \(\ell\in\{1,\dots,N\}\), if \(\bar \delta_h(\ell)>1\), then \(\bar {\mathcal Y}_h(t_0+\ell)\cap \hat {\mathcal Y}_h(t_0+\ell)=\emptyset.\)
\end{proposition}

From Proposition~\ref{prop:surrogate_sep}, the offline exposure budget problem can be converted into a condition for making \(\bar\delta_h(\ell)\) exceed \(1\) within the prescribed horizon. The next theorem provides a necessary lower threshold on the exposure budget for certifying predicted separation for all attack hypotheses.

\begin{theorem}
\label{thm:offline_budget_lower}
Consider the surrogate separation tendency in~\eqref{eq: surrogate separation tendency definition} under Assumption~\ref{ass:output_visibility}. 
Given the attack hypothesis set $\mathcal{H}$,
for each \(h\in \mathcal{H}\), define
\[
L_h=\{\ell\in\{1,\dots,N\}: C_hM_\ell\neq 0\}.
\]
Let
\begin{equation}
\bar u_{\min}:=
\max_{h\in \mathcal{H}}\;
\min_{\ell\in L_h}
\frac{\big(\rho_h(\ell)-\|e_h^0(\ell)\|_\infty\big)_+}
{\|C_hM_\ell\|_\infty}.
\label{eq:26}
\end{equation}
If \(\bar u<\bar u_{\min}\), then there exists at least one
hypothesis \(h^\star\in \mathcal{H}\) such that
\begin{equation}\label{eq: theorem2 conclusion}
    \bar \delta_{h^\star}(\ell)\le 1,\qquad \forall \ell\in L_{h^\star}.
\end{equation}
Therefore, the predicted set separation cannot be certified.
\end{theorem}

The next result gives a sufficient threshold for the exposure budget.

\begin{theorem}
\label{thm:offline_budget_upper}
Consider the surrogate separation tendency in~\eqref{eq: surrogate separation tendency definition} under Assumption~\ref{ass:output_visibility}.
Given the attack hypothesis set $\mathcal{H}$,
suppose that there exists a step
\(\ell^\star\in\{1,\dots,N\}\) and a normalized vector
\(r_{\ell^\star}\in\mathbb R^{\ell^\star n_u}\) such that
\[
\|r_{\ell^\star}\|_\infty\le 1,\quad
\beta_h=\|C_hM_{\ell^\star}r_{\ell^\star}\|_\infty>0,
\quad \forall h\in \mathcal{H}.
\]
Let
\begin{equation}
\bar u^{\mathrm{suf}}:=
\max_{h\in \mathcal{H}}
\frac{\rho_h(\ell^\star)+\|e_h^0(\ell^\star)\|_\infty}
{\beta_h}.
\label{eq:ubar_suf}
\end{equation}
If \(\bar u>\bar u^{\mathrm{suf}}\), then by choosing \(d_{\ell^\star}=\bar u\, r_{\ell^\star},\)
one has
\begin{equation}\label{eq: theorem3 conclusion}
   \bar  \delta_h(\ell^\star)>1.
\end{equation}
\end{theorem}

Theorem~\ref{thm:offline_budget_lower} and Theorem~\ref{thm:offline_budget_upper} provide a budget guidance for the selection of \(\bar u\). 
In particular, if \(\bar u<\bar u_{\min}\), then the offline surrogate separation cannot be certified for all  attack hypotheses; 
if \(\bar u>\bar u^{\mathrm{suf}}\), then there exists an excitation direction that guarantees predicted separation for all attack hypotheses. 
Since the actual secure state set \(\mathcal{X}_\mathcal{S}(k)\) still depends on online secure-sensor measurements, these results are used as offline guidance for budget selection, while the actual exposure inputs are generated online by \eqref{eq: opt1}--\eqref{eq: opt4}.
The entire exposure framework is shown in Algorithm 1.

\begin{algorithm}[t]
\caption{Exposure Framework for Stealthy Attacks}
\label{alg:exposure_framework_compact}
\begin{algorithmic}[1]
\Require
Exposure start time $t_0$; horizon $N$; suspicious sensor set $\mathcal A$; secure sensor set $\mathcal S$; initial secure state set $\mathcal X_{\mathcal S}(t_0)$; initial hypothesis set $\mathcal H(t_0)=\{1,\dots,2^{|\mathcal A|}\}$; weights $\{w_h\}_{h\in\mathcal H(t_0)}$; increment parameter $\varepsilon>0$.
\Ensure
Detected attacked sensor set $\mathcal D$.

\State \textbf{Offline:} Compute $\bar u_{\min}$ and $\bar u^{\mathrm{suf}}$ by Theorems~\ref{thm:offline_budget_lower}--\ref{thm:offline_budget_upper}; choose $\bar u$
\State $k\gets t_0$, $\mathcal D\gets\emptyset$
\While{$k<t_0+N$}
    \State Construct $\mathcal X_{\mathcal S}(k)$ by Lemma~\ref{lemma: secure state set}
    \ForAll{$i\in\mathcal A\setminus\mathcal D$}
        \State \(\mathcal Y_i(k)\gets C_i\mathcal X_{\mathcal S}(k)\oplus\mathcal V_i\)
    \EndFor
    \State $\mathcal D_{\rm new}\gets\{\,i\in\mathcal A\setminus\mathcal D: y_i(k)\notin \mathcal Y_i(k)\,\}$
    \State $\mathcal D\gets\mathcal D\cup\mathcal D_{\rm new}$
    \State $\mathcal H(k)\gets\{\,h\in\mathcal H(k): \mathcal D_{\rm new}\subseteq\mathcal F(h)\,\}$
    \ForAll{$h\in\mathcal H(k)$}
        \State Construct $\mathcal Y_h(k), \hat{\mathcal Y}_h(k)$
        \State $\delta_h(k)\gets\hat\delta(\mathcal Y_h(k),\hat{\mathcal Y}_h(k))$
    \EndFor
    \If{$(\mathcal H(k)=\{h^*\} \land \delta_{h^\star}(k)>1$ \textbf{or} $(\delta_h(k)>1,\ \forall h\in\mathcal H(k))$}
        \State \textbf{break}
    \EndIf
    \State Solve \eqref{eq: opt1}--\eqref{eq: opt4} to obtain $d(k)$, apply $u^s(k)=u^\star(k)+d(k)$
    \State $k\gets k+1$
\EndWhile
\State $k_{\rm stop}\gets k$; \Return $\mathcal D$
\end{algorithmic}
\end{algorithm}

\section{CASE STUDIES}\label{sec: 5}
To demonstrate the effectiveness of the proposed framework, we conduct simulations on a UAV navigation system equipped with LiDAR, GNSS, IMU, and a barometer, under simultaneous attacks on GNSS and LiDAR.

\subsection{UAV Model}
A discrete-time 3D UAV navigation model with state \(x_k=[p_x,p_y,p_z,v_x,v_y,v_z]^\top\) and input \(u_k=[a_x,a_y,a_z]^\top\) is considered, with sampling period \(dt=0.1\). The system matrices are given by
\[
A = \begin{bmatrix}
    I_3 & dt I_3 \\
    0 & I_3\\
\end{bmatrix}, \quad 
B =  \begin{bmatrix}
\frac{1}{2}dt^2 I_3\\
dt I_3\\
\end{bmatrix}.
\]
The measurement matrices are $C_{Lidar}=[I_3\; 0]$, $C_{gnss}=[I_6]$, $C_{imu}=[0 \; I_3]$, and $C_{baro} = [0\;0\;1\;0\;0 \; 0]$.
The process noise and measurement noise standard deviations are set to $w_k \in [-0.02, 0.02]$, $\sigma_{imu}=0.08$ m/s, $\sigma_{baro}=0.6$ m, $\sigma_{gnss,p}=0.6$ m, $\sigma_{gnss,v}=0.15$ m/s, and $\sigma_{Lidar}=0.9$ m.

A standard Kalman filter with a \(\chi^2\) detector is adopted for sensor fusion and anomaly detection. The thresholds for LiDAR and GNSS are $7.81$ and $12.59$, respectively, with the significance level $0.05$.
The reference trajectory is designed as a smooth helical path. Accordingly, the desired position is defined as \(p_x^{r}(t) = R\cos(\omega t), p_y^{r}(t) = R\sin(\omega t), p_z^{r}(t) = z_0 + A_z\sin(\omega_z t),\)
where $R=80$ m, $\omega=0.04$ rad/s, $z_0=50$ m, $A_z=10$ m, and $\omega_z=0.08$ rad/s.



\subsection{Stealthy Deception Attack}
\begin{figure}
    \centering
    \includegraphics[width=0.85\linewidth]{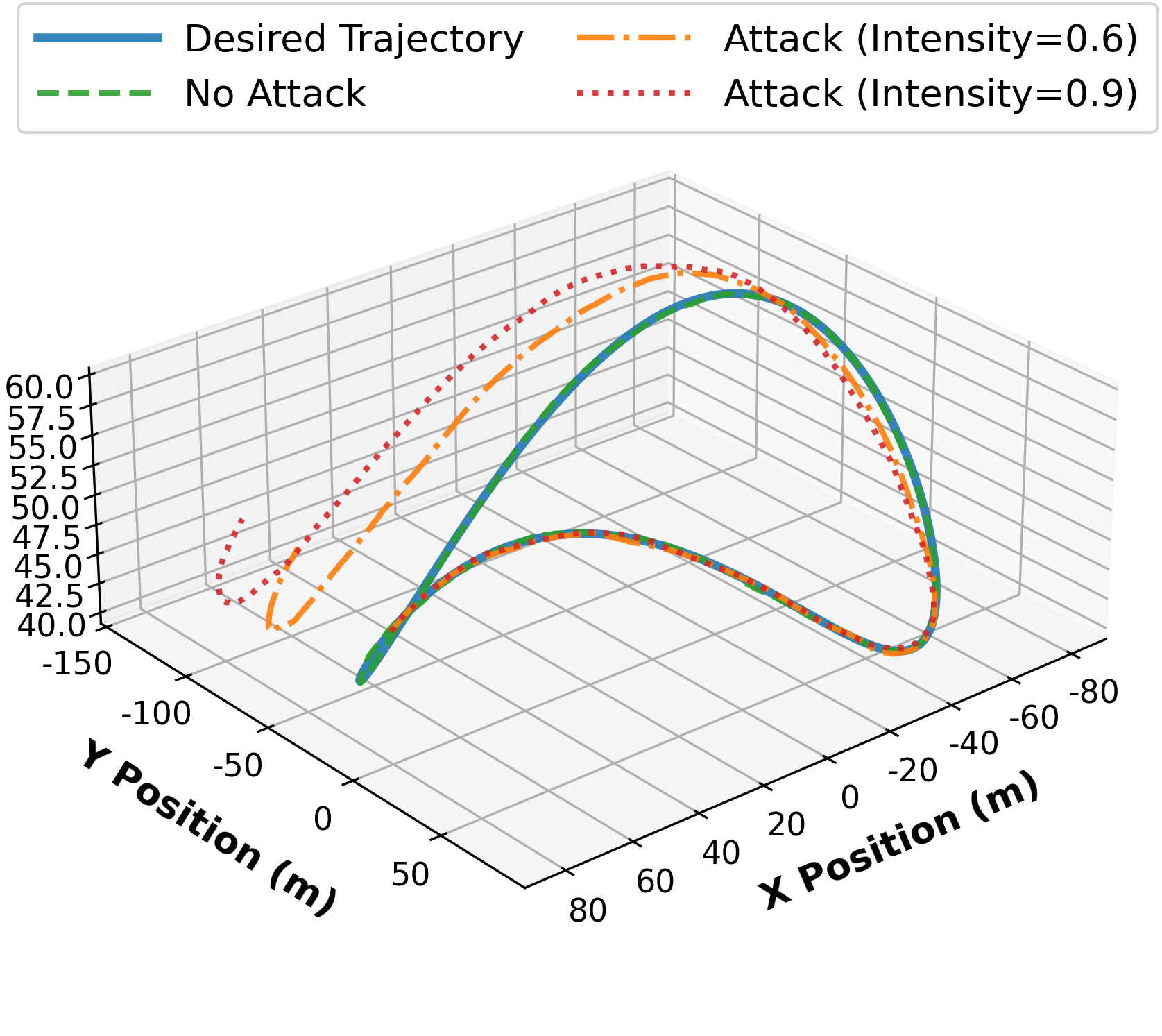}
    \caption{3D Trajectories of UAV Under Different Attack Scenarios.}
    \label{fig: 3D Trajectories}
\end{figure}

We consider two stealthy attack intensities, 0.6 and 0.9, where larger intensity corresponds to stronger measurement tampering while remaining undetected by the \(\chi^2\) detector. In both cases, the attack is launched within \(60-160\) s and causes a visible trajectory deviation without triggering alarms as shown in Fig. \ref{fig: 3D Trajectories}.

\subsection{Stealthy Attack Exposure}

At \(60\) s, the detector outputs indicate suspicious behavior, and three attack hypotheses are considered: GNSS only, LiDAR only, and simultaneous GNSS–LiDAR attack. 
The exposure horizon is set to $50$ steps.
The offline guidance gives \(\bar u_{\min}=1.47\) and \(\bar u^{\mathrm{suf}}=2.94\), and we choose \(\bar u=2\). 
The optimization weights are chosen as $[1,5,1]$.

\begin{figure}
    \centering
    \includegraphics[width=0.85\linewidth]{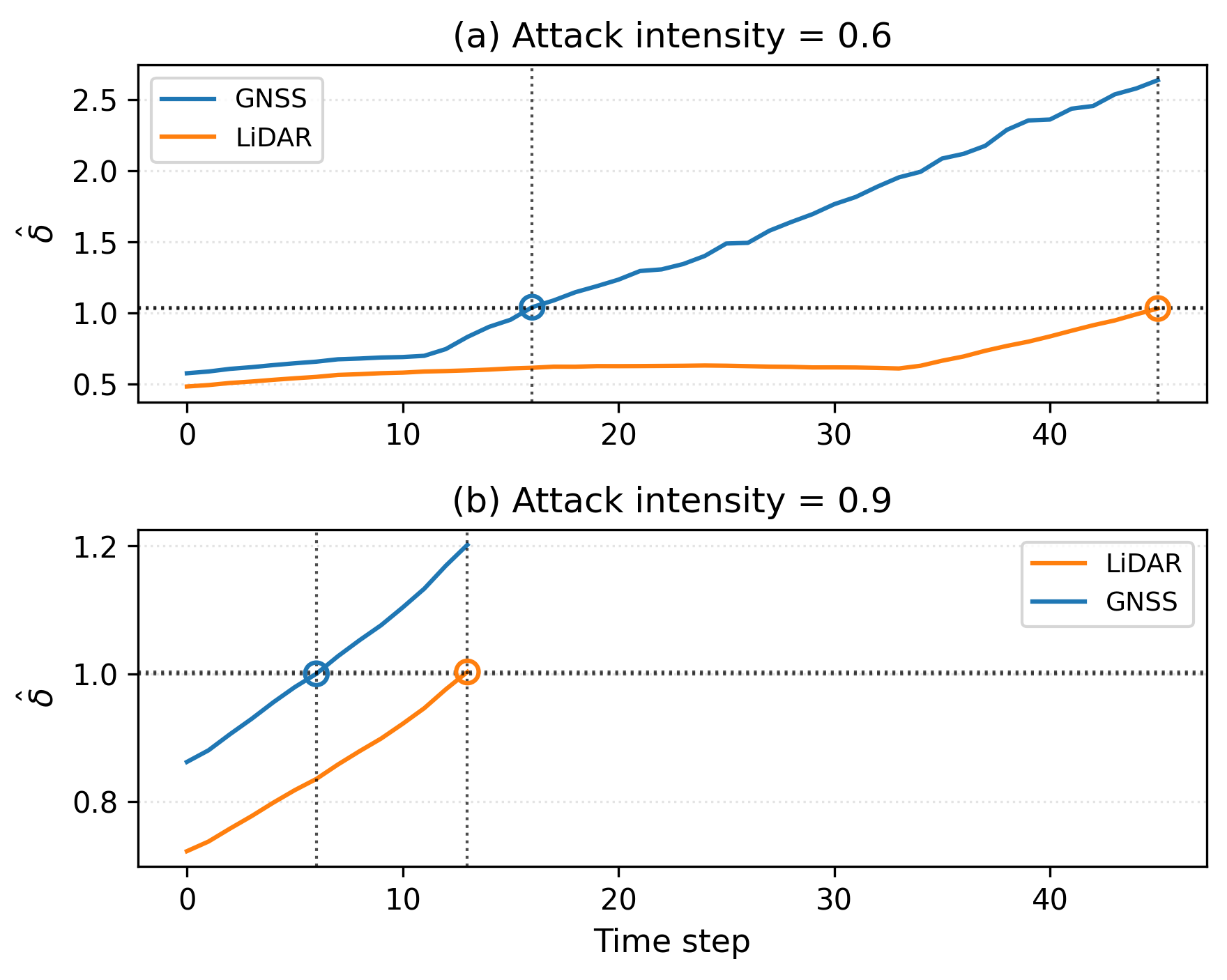}
    \caption{Time evolution of the separation tendency for GNSS and LiDAR during the exposure phase. The hollow circles mark the first time step at which the corresponding sensor is detected as attacked.}
    \label{fig: separation tendency}
\end{figure}

We represent the GNSS and LiDAR measurements in zonotopic form.
Fig.~\ref{fig: separation tendency} depicts the time evolution of the corresponding separation tendency \(\hat{\delta}\) between each measurement zonotope and the admissible secure set during the exposure phase.
Under attack intensity $0.6$, GNSS and LiDAR are first exposed at time steps $17$ and $46$, respectively, while under attack intensity 0.9 they are exposed earlier, at time steps $7$ and $14$.
These results also indicate that a higher attack intensity leads to a faster increase in the separation tendency, thereby enabling earlier attack exposure.

\section{Conclusion}\label{sec: 6}
This paper investigated the active exposure of stealthy deception attacks in sensor-fusion uncertain systems. 
By constructing defender-side admissible output sets and hypothesis-dependent attack output sets, we developed a receding-horizon exposure framework that enlarges their separation via bounded auxiliary inputs.
A sufficient detection condition was established to connect set separation with attack detectability, and an offline guidance was further derived to support budget selection.
Simulation results on a UAV navigation system under GNSS and LiDAR attacks validated the effectiveness of the proposed method.
Future work will extend to other types of cyber attacks.

\appendix
\subsection{Proof of Theorem \ref{thm:detection_condition}}
By construction, \(\mathcal X_{\mathcal S}(k)\) over-approximates all states
consistent with the defender dynamics, the secure-sensor measurements, and the
bounded uncertainties. Hence, if sensor \(i\) is attack-free, then there exist
\(x(k)\in\mathcal X_{\mathcal S}(k)\) and \(v_i(k)\in\mathcal V_i\) such that
\(
y_i(k)=C_i x(k)+v_i(k),
\)
which implies \(y_i(k)\in \mathcal{Y}_i(k)\). 
Therefore,
\(y_i(k)\notin \mathcal Y_i(k)\) indicates that the received measurement is
inconsistent with all attack-free outputs compatible with the secure state set,
and thus sensor \(i\) is detected as attacked.

\subsection{Proof of Proposition \ref{prop:surrogate_sep}}
Suppose that there exists some $h\in\mathcal H$ such that 
\(
\bar{\mathcal{Y}}_h(t_0+j^\star)\cap \hat{\mathcal{Y}}_h(t_0+j^\star)\neq\emptyset.
\)
By the zonotopic representations
\begin{align*}
    \bar{\mathcal Y}_h(t_0+j^\star)
=
\left\langle
\bar c_h(j^\star),\,
\big[\,C_h\bar H_{\mathcal S}(t_0+j^\star)\;\; H_{h,v}\,\big]
\right\rangle,\\
\hat{\mathcal Y}_h(t_0+j^\star)
=
\left\langle
\hat c_h(j^\star),\,
\big[\,C_h H_h(t_0+j^\star)\;\; H_{h,v}\,\big]
\right\rangle,
\end{align*}

there exist vectors
$\xi_1,\xi_2,\zeta_1,\zeta_2$ satisfying
\[
\|\xi_1\|_\infty\le 1,\quad
\|\xi_2\|_\infty\le 1,\quad
\|\zeta_1\|_\infty\le 1,\quad
\|\zeta_2\|_\infty\le 1
\]
such that
\begin{equation}
\begin{split}
\bar c_h(j^\star)
+ C_h\bar H_{\mathcal S}&(t_0+j^\star)\xi_1
+ H_{h,v}\zeta_1
=\\
&\hat c_h(j^\star)
+ C_h H_h(t_0+j^\star)\xi_2
+ H_{h,v}\zeta_2.
\end{split}\nonumber
\end{equation}
Define
\[
e_h(j^\star)
=
\mathcal{C}_hH_h(t_0+j^\star)\xi_2
-
\mathcal{C}_h\bar H_\mathcal{S}(t_0+j^\star)\xi_1
+
H_{h,v}(\zeta_2-\zeta_1).
\]
Taking the infinity norm on both sides gives
\begin{equation}
    \begin{split}
       & \|e_h(j^\star)\|_\infty
\le \\
&\left\|
\left[
C_h\bar H_\mathcal{S}(t_0+j^\star)\;\;
C_hH_h(t_0+j^\star)\;\;
H_{h,v}\;\;
H_{h,v}
\right]
\right\|_\infty,
    \end{split}
\end{equation}
Therefore, \(\|\bar c_h(j^\star)-\hat c_h(j^\star)\|_\infty
\le
\rho_h(j^\star)\) holds,
which contradicts \(\bar \delta_h(j^\star)>1\) by eq.~\eqref{eq: surrogate separation tendency definition}.

\subsection{Proof of Theorem \ref{thm:offline_budget_lower}}
By \eqref{eq:25}, we obtain \(e_h  (j)=e_h^{0}(j)+\mathcal{C}_hM_j d_{j}.\)
Taking the infinity norm on both sides gives the following:
\begin{align}
    \|e_h(j)\|_\infty  \nonumber
&\le \|e_h^0(j)\|_\infty+\|C_hM_j d_{j}\|_\infty  \nonumber\\
&\le \|e_h^0(j)\|_\infty+\|C_hM_j\|_\infty\,\|d_{j}\|_\infty \nonumber\\
&\le \|e_h^0(j)\|_\infty+\|C_hM_j\|_\infty\,\bar u. \label{eq: proof theorem2-1}
\end{align}
Now let 
\(
h^\star\in
\arg\max_{h\in \mathcal{H}}
\min_{j\in \mathcal{L}_{h}}
\frac{\big(\rho_h(j)-\|e_h^{0}(j)\|_\infty\big)_+}
{\|\mathcal C_h M_j\|_\infty}.
\)
If $\bar u<\bar u_{\min}$, for hypothesis $h^\star$ and every $j\in\mathcal J_{h^\star}$, we obtain
\[
\bar u<
\frac{\big(\rho_{h^\star} (j)-\|e_{h^\star}^{0}(j)\|_\infty\big)_+}
{\|\mathcal C_{h^\star} M_j\|_\infty}.
\]
Therefore, we have 
\(
\|e_{h^\star}^{0}(j)\|_\infty
+\|\mathcal C_{h^\star} M_j\|_\infty \bar u
<
\rho_{h^\star}(j).
\)
Combining this with \eqref{eq: proof theorem2-1} yields
\(
\|e_{h^\star}(j)\|_\infty<\rho_{h^\star}(j),
\quad \forall j\in\mathcal J_{h^\star},
\)
which implies~\eqref{eq: theorem2 conclusion} by \eqref{eq: surrogate separation tendency definition}.

\subsection{Proof of Theorem \ref{thm:offline_budget_upper}}
Let \(d_{j^\star} = \bar u\,r_{j^\star}\).
Since $\|r_{j^\star}\|_\infty\le 1$, the input sequence
$d_{j^\star}$ satisfies the budget constraint
$\|d(k)\|_\infty\le \bar u$ for
$k=t_0,\dots,t_0+j^\star-1$.

Repeated expansion of $\bar c_s(j)$ up to $j^\star$ yields
\begin{equation}
e_h(j^\star)
=
e_h^{0}(j^\star)+C_hM_{j^\star}d_{j^\star}.\nonumber
\label{eq:ey_star}
\end{equation}
Using the reverse triangle inequality and
$d_{j^\star}=\bar u\,r_{j^\star}$ gives
\begin{align}
\|e_h(j^\star)\|_\infty
&\ge
\|C_hM_{j^\star}d_{j^\star}\|_\infty
-\|e_h^{0}(j^\star)\|_\infty \nonumber\\
&=
\bar u\,\|C_hM_{j^\star}r_{j^\star}\|_\infty
-\|e_h^{0}(j^\star)\|_\infty \nonumber\\
&=
\bar u\,\beta_h-\|e_h^{0}(j^\star)\|_\infty .
\label{eq:ey_lower}
\end{align}
If $\bar u>\bar u^{\mathrm{suf}}$, then by
\eqref{eq:ubar_suf}, we obtain
\(
\bar u\,\beta_h-\|e_h^{0}(j^\star)\|_\infty\
>
\rho_h(j^\star).
\)
Combining this with \eqref{eq:ey_lower} yields
\[
\|e_h(j^\star)\|_\infty>\rho_h(j^\star),
\qquad \forall h\in\mathcal H,
\]
which implies~\eqref{eq: theorem3 conclusion} by \eqref{eq: surrogate separation tendency definition}.

\end{document}